\documentclass[aps,prd,twocolumn,showpacs,superscriptaddress,groupedaddress,nofootinbib]{revtex4}  

\usepackage{amsmath,amssymb,bm,color,comment,dcolumn,mathrsfs,tabularx,graphicx,Macro}

\def\planck{{\it Planck} }
\def\BK{{\sc Bicep} / {\it Keck Array} }
\def\LB{LiteBIRD}

\def\bn{\bm{\nabla}}
\def\hatn{\hat{\bm{n}}}
\def\l{\ell}
\def\grad{\phi}
\def\hx{\widehat{x}}

\def\hE{\widehat{E}}
\def\hC{\widehat{C}}
\def\Cgg{C^{\grad\grad}}
\def\hCEE{\hC^{\rm EE}}

\begin{document}

\title{Delensing Cosmic Microwave Background B-modes with the Square Kilometre Array Radio Continuum Survey}

\author{Toshiya Namikawa}
\affiliation{Department of Physics, Stanford University, Stanford, CA 94305, USA}
\affiliation{Kavli Institute for Particle Astrophysics and Cosmology, SLAC National Accelerator
Laboratory, Menlo Park, CA 94025, USA}
\author{Daisuke Yamauchi}
\affiliation{Research Center for the Early Universe, Graduate School of Science, 
The University of Tokyo, Bunkyo-ku, Tokyo 113-0033, Japan}
\author{Blake Sherwin}
\affiliation{Department of Physics, University of California, Berkeley, CA, 94720, USA}
\author{Ryo Nagata}
\affiliation{High Energy Accelerator Research Organization (KEK), Tsukuba, Ibaraki 305-0801, Japan}

\date{\today}

\begin{abstract}
We explore the potential use of the Radio Continuum (RC) survey conducted by the Square Kilometre 
Array (SKA) to remove (delens) the lensing-induced B-mode polarization and thus enhance future cosmic 
microwave background (CMB) searches for inflationary gravitational waves. Measurements of large-scale 
B-modes of the CMB are considered to be the best method for probing gravitational waves from the 
cosmic inflation. Future CMB experiments will, however, suffer from contamination by non-primordial 
B-modes, one source of which is the lensing B-modes. Delensing, therefore, will be required for further 
improvement of the detection sensitivity for gravitational waves. Analyzing the use of the 
two-dimensional map of galaxy distribution provided by the SKA RC survey as a lensing mass tracer, we 
find that joint delensing using near-future CMB experiments and the SKA phase 1 will improve 
the constraints on the tensor-to-scalar ratio by more than a factor of $\sim 2$ compared to those 
without the delensing analysis. Compared to the use of CMB data alone, the inclusion of 
the SKA phase 1 data will increase the significance of the constraints on the tensor-to-scalar ratio 
by a factor $1.2$--$1.6$. For \LB combined with a ground-based experiment such as Simons Array and 
Advanced ACT, the constraint on the tensor-to-scalar ratio when adding SKA phase 2 data is improved 
by a factor of $2.3$--$2.7$, whereas delensing with CMB data alone improves the constraints by only 
a factor $1.3$--$1.7$. We conclude that the use of SKA data is a promising method for delensing 
upcoming CMB experiments such as \LB. 
\end{abstract}

\maketitle


\section{Introduction} \label{sec.1}

Measurements of the B-mode polarization of the cosmic microwave background (CMB) on angular scales 
larger than a few dozen arc-minutes have long been considered as the best method to probe the 
primordial gravitational waves \cite{Polnarev:1985,Seljak:1996gy,Seljak:1997,Kamionkowski:1997}. The 
recent BICEP/Keck Array observation reported upper bounds on the tensor-to-scalar ratio, $r_{0.05}<0.09$ and 
$r_{0.05}<0.07$ \footnote{The subscript $0.05$ is the pivot scale of the primordial power spectrum in 
units of Mpc$^{-1}$.}, at $95$\% C.L. using B-modes alone and combining the B-mode results with \planck 
temperature analysis, respectively \cite{BKVI}. The detection of the B-mode signals induced by the 
primordial gravitational waves is one of the main targets in many ongoing and future CMB experiments.

On large scales, however, other secondary B-modes produced by Galactic foreground emission and 
gravitational lensing are expected to dominate over the B-mode signals from the primordial 
gravitational waves (the primary B-modes). Many studies have been devoted to foreground rejection 
techniques (e.g., \cite{Dunkley:2008am,Betoule:2009,Katayama:2011eh}) and it appears possible to 
remove the foreground contamination sufficiently to detect primordial gravitational waves at 
the level of $r\sim 0.001$ \cite{Katayama:2011eh}. The contribution of the lensing B-modes can be 
estimated as a convolution between the observed E-modes and the CMB lensing-mass map or any 
lensing-mass tracers which significantly correlate with the CMB lensing mass map
\cite{Kesden:2002,Knox:2002,Seljak:2003pn,Marian:2007sr,Smith:2010gu,Simard:2014aqa,Sherwin:2015baa}.
Indeed, the lensing B-modes have been recently estimated from the precise measurements of the CMB 
lensing mass map and cosmic infrared background (CIB)
\cite{Hanson:2013daa,vanEngelen:2014zlh,Ade:2015zua}. Subtraction of the estimated lensing B-modes 
from the observed B-modes, usually referred to as {\it delensing}, will improve detection sensitivity 
for the primary B-modes, and will be required for ongoing and future CMB experiments.

Future CMB experiments such as significant upgrade of \BK \cite{BICEP3} and \LB \cite{LiteBIRD}
will have high sensitivity to the large-scale B-mode polarization. However, they will observe with 
large beams, so that internal delensing (using their data alone to recover the lensing signal) is 
not effective. Efficient delensing can only be achieved by the use of external data sets.

In this paper, we explore the potential use of the Square Kilometre Array (SKA) data for delensing. 
Previous studies of the delensing analysis with future radio surveys have assumed the use of an HI 
($21$cm-line) intensity mapping survey to reconstruct the lensing mass map at high 
redshift~\cite{Sigurdson:2005cp}, or ellipticity measurements of each galaxy to extract lensing 
information \cite{Marian:2007sr} (see also \cite{Brown:2015ucq} for a review of the SKA weak 
lensing measurement). These delensing techniques are, however, not efficient unless we can measure 
the sources at very high-redshifts $z>10$ where the foreground uncertainties are significant. 
Instead of measuring the lensing effect from the HI intensity map or shapes of each galaxy, we use 
observables from the Radio Continuum (RC) survey conducted by the SKA, and apply them to the 
delensing analysis in a similar way of the CIB delensing proposed recently by 
Refs.~\cite{Simard:2014aqa,Sherwin:2015baa}. Galaxies identified as radio sources through the SKA RC 
survey are located at higher redshifts and their number density is sufficient to be comparable to 
that in forthcoming optical surveys. The SKA RC survey, therefore, provides a two-dimensional 
integrated-mass map at high redshifts whose gravitational potential induces most of the CMB lensing 
signal.

This paper is organized as follows: In Sec.~\ref{method}, we describe our method to evaluate 
delensing performance. In Sec.~\ref{results}, we show results of the expected efficiency of the 
delensing analysis for several cases of the experimental specifications including \LB. Then we 
show the effects of the uncertainties in the bias model and distribution functions on the results. 
We also discuss the comparison between our results and the CIB delensing. Sec.~\ref{summary} is 
devoted to a discussion and to our conclusions.

Throughout this paper, we assume a flat $\Lambda$CDM model characterized by six parameters. 
The cosmological parameters have the best-fit values of Planck 2015 results \cite{Ade:2015xua}.

\section{Delensing with lensing-mass tracers} \label{method}

Here we briefly summarize our methods to evaluate expected delensing performance using lensing-mass tracers.

\subsection{Lensing B-modes}

We denote the primary polarization anisotropies as $Q\pm\iu U$. The lensed polarization anisotropies 
observed in the direction $\hatn$, are given by (e.g., \cite{Zaldarriaga:1998ar}):
\al{
	[Q^{\rm lens}\pm\iu U^{\rm lens}](\hatn) &= [Q\pm\iu U](\hatn + \bn\grad(\hatn)) 
	\,, \label{Eq:remap}
}
where $\grad$ is the CMB lensing potential. Instead of being expressed as spin-$2$ quantities, the 
following E- and B-mode polarizations are useful to analyse the polarization anisotropies in harmonic 
space (e.g., \cite{Zaldarriaga:1998ar}):
\al{
	[E \pm \iu B ]_{\l m} = -\Int{}{\hatn}{} {}_{\pm 2} Y_{\l m}^*(\hatn) [Q\pm \iu U](\hatn)  \,, 
}
where we denote the spin-$2$ spherical harmonics as ${}_{\pm 2}Y_{\l m}$. Similarly, with the spin-$0$ 
spherical harmonics, $Y_{\l m}$, the CMB lensing potential is transformed into the harmonic space as
\al{
	\grad_{LM} = \Int{}{\hatn}{} Y_{LM}^*(\hatn) \grad (\hatn)  \,.
}
Expanding Eq.~\eqref{Eq:remap} up to the first order of the CMB lensing potential, the B-modes are 
described as (e.g., \cite{Hu:2000ee}) 
\al{
	B^{\rm lens}_{\l m} = \mS{B}_{\l m}^{\l'm'LM} E_{\l'm'}\grad_{LM}
	\,, \label{Eq:Lensing-E-to-B}
}
where we ignore the primary B-mode and simplify the above equation by defining a convolution operator 
for two multipole moments:
\al{
	\mS{B}_{\l m}^{\l'm'LM} \equiv -\iu\sum_{\l'm'}\sum_{LM} 
		\Wjm{\l}{\l'}{L}{m}{m'}{M}\mC{S}_{\l\l'L}
	\,. 
}
The quantity $\mC{S}_{\l\l'L}$ represents the mode coupling induced by the lensing: 
\al{ 
	\mC{S}_{\l\l'L} 
		&= \sqrt{\frac{(2\l+1)(2\l'+1)(2L+1)}{16\pi}}\Wjm{\l}{\l'}{L}{2}{-2}{0}
	\notag \\
		&\qquad\times [-\l(\l+1)+\l'(\l'+1)+L(L+1)] 
	\,. \label{Eq:Spm}
} 
Here the above quantity is unity if $\l+\l'+L$ is an odd integer and zero otherwise.

\subsection{Residual B-modes}

Delensing of the B-modes with lensing-mass tracers has been discussed in 
Refs.~\cite{Marian:2007sr,Smith:2010gu,Simard:2014aqa,Sherwin:2015baa}.
The delensed (residual) B-modes are given in the following from: 
\al{
	B^{\rm res}_{\l m} = \mS{B}^{\l'm'LM}_{\l m} \left[E_{\l'm'}\grad_{LM}
		- W_{\l'}\hE_{\l'm'}\sum_i a^i_{\l\l'L} \hx^i_{LM}\right]
	\,, \label{Eq:delens-B}
}
where $\hE_{\l'm'}$ is the observed E-modes including the noise contribution, 
$\hx^i_{LM}=x^i_{LM}+n^i_{LM}$ is an $i$-th observed mass tracer which correlates with the CMB lensing 
potential. The quantity $W_\l$ is the E-mode Wiener filter defined as $W_\l=C_\l^{\rm EE}/\hCEE_\l$,
where $C_\l^{\rm EE}$ and $\hCEE_\l$ denote the angular power spectra of $E$ and $\hE$, respectively.

The coefficients $a^i_{l\l'L}$ are simply determined so that the variance of the residual B-modes is 
minimized as follows. From Eq.~\eqref{Eq:delens-B}, the angular power spectrum of the residual 
B-modes is given by
\al{
	C^{\rm BB,res}_\l 
		&= \Xi^{\l'L}_\l \bigg[ C^{\rm EE}_{\l'}\Cgg_L 
			- 2 W_{\l'}C^{\rm EE}_{\l'}\sum_i a^i_{\l\l'L}C^{\grad x^i}_L
	\notag \\
		&+ W_{\l'}C^{\rm EE}_{\l'} \sum_{i,j} a^i_{\l\l'L}a^j_{\l\l'L}\hC^{x^ix^j}_L\bigg]
	\,. \label{Eq:Bres-var}
}
Here the measured cross-power spectrum between $x^i$ and $x^j$ ($\hC^{x^ix^j}_L$) is the sum of 
the signal ($C^{x^ix^j}_L$) and noise ($N^{x^ix^j}_L$). The operator $\Xi^{\l'L}_\l$ is defined as
\al{
	\Xi^{\l'L}_\l \equiv |\mS{B}^{\l'm'LM}_{\l m}|^2  \,.
}
The above operator is independent of the integers $m$, $m'$ and $M$ due to the properties of the 
Wigner 3j symbols (see e.g., Refs.~\cite{QTAM,Hu:2000ee,Namikawa:2011cs}). The coefficients 
$a^i_{\l\l'L}$ which minimize the power spectrum are the solution of the following equation: 
\al{
	0 = \PD{C^{\rm BB,res}_{\l}}{a^i_{\l\l'L}} 
		&\propto -2 C^{\grad x^i}_L + 2\sum_j a^j_{\l\l'L} \hC^{x^ix^j}_L
	\,,\label{eq:derivatives}
}
For simplicity of notation, we introduce the tensor expression of the variables; 
$\{\bm{v}_L\}_i= C^{\grad x^i}_L$, $\{\bR{C}_L\}_{ij} = \hC^{x^ix^j}_L$ 
and $\{\bm{a}_L\}_i = a^i_{\l\l'L}$. With these notations, Eq.~\eqref{eq:derivatives} is recast as
\al{
	0 = \PD{C^{\rm BB,res}_{\l}}{{\bm a}_L} \propto - 2\left(\bm{v}_L - \bR{C}_L\bm{a}_L\right)
	\,.
}
The solution is then simply expressed as
\al{
	\bm{a}_L = \bR{C}_L^{-1}\bm{v}_L  \,. \label{Eq:filter}
}
Note that, substituting Eq.~\eqref{Eq:filter} into Eq.~\eqref{Eq:Bres-var}, 
the residual B-mode power spectrum is given by
\al{
	C^{\rm BB,res}_\l 
		&= \Xi^{\l'L}_\l C^{\rm EE}_{\l'}\Bigl[\Cgg_L 
			- W_{\l'}\left(2\bm{a}_L\cdot\bm{v}_L - (\bm{a}_L)^t \bR{C}_L \bm{a}_L\right)\Bigr]
	\notag \\
		&= \Xi^{\l'L}_\l C^{\rm EE}_{\l'}\left(\Cgg_L - W_{\l'}\bm{a}_L\cdot\bm{v}_L\right)
	\,. \label{Eq:delensed-ClBB}
}

Now we turn to discuss specific cases. Hereafter we consider the radio source distribution via 
the RC survey conducted by the SKA as one of promising candidates of the mass tracers.
The source number density observed via the RC survey is projected onto a two-dimensional map, $I(\hatn)$. 
If we only use the $I$ map ($x^1=I$), the residual B-mode power spectrum becomes 
\cite{Simard:2014aqa,Sherwin:2015baa} 
\al{
	C^{\rm BB,res}_\l &= \Xi_\l^{\l'L} C^{\rm EE}_{\l'}\Cgg_L \left(1-W_{\l'}\rho^2_L\right)
	\,, \label{Eq:resBB:CIB}
}
where $\rho_L$ is a correlation coefficient:
\al{
	\rho^2_L = \frac{(C_L^{\grad I})^2}{\hC_L^{II}\Cgg_L}  \,. \label{Eq:corr}
}
As the amplitude of the correlation coefficient increases, the residual B-mode power spectrum decreases. 
The correlation coefficient becomes small if the noise and residual foreground of the mass tracer 
become large compared to the signal power spectrum. 

We can also use both the CMB lensing potential and SKA data ($x^1=\grad$ and $x^2=I$) for 
the delensing analysis. In this case, the covariance is given by
\al{
	\bR{C}_L &= \begin{pmatrix} \Cgg_L+N_L^{\grad\grad} & C_L^{\grad I} 
			\\ C_L^{\grad I} & C_L^{II} + N_L^{II} \end{pmatrix}
	\,. \label{Eq:cov}
}
The coefficients of Eq.~\eqref{Eq:filter} are then described as
\al{
	\binom{a_L^{\grad}}{a^I_L}
	&= \begin{pmatrix} \Cgg_L+N_L^{\grad\grad} & C_L^{I\grad} 
			\\ C_L^{I\grad} & C_L^{II} + N_L^{II} \end{pmatrix}^{-1} \binom{\Cgg_L}{C_L^{I\grad}}
	\notag \\
	&= \frac{1}{1-\beta_L\rho^2_L}\binom{\beta_L(1-\rho^2_L)}{\gamma_L(1-\beta_L)}
	\,, \label{Eq:filter-comb}
}
where we define
\al{
	\beta_L  &= \frac{\Cgg_L}{\Cgg_L+N_L^{\grad\grad}}  \,, \label{Eq:beta} \\
	\gamma_L &= \frac{C_L^{I\grad}}{C_L^{II}+N_L^{II}}  \,. \label{Eq:gamma}
}
Substituting Eqs.~\eqref{Eq:filter-comb} into Eq.~\eqref{Eq:delensed-ClBB}, the residual B-mode 
power spectrum becomes
\al{
	&C^{\rm BB,res}_\l = \Xi^{\l'L}_{\l} C^{\rm EE}_{\l'}
		\bigg[\Cgg_L - \frac{W_{\l'}}{1-\beta_L\rho_L^2}
	\notag \\
	&\qquad\qquad \times \left(\beta_L(1-\rho^2_L)\Cgg_L+\gamma_L(1-\beta_L)C_L^{\grad I}\right)\bigg]
	\notag \\
	&\qquad = \Xi^{\l'L}_\l C^{\rm EE}_{\l'} \Cgg_L\bigg[1-\frac{W_{\l'}}{1-\beta_L\rho_L^2}
	\notag \\
	&\qquad\qquad \times \left(\beta_L(1-\rho^2_L)+\rho^2_L(1-\beta_L)\right)\bigg]
	\,. \label{Eq:delensed-comb}
}
In our analysis, the reconstruction noise of the CMB lensing potential, $N_L^{\grad\grad}$, is 
computed from the iterative method \cite{Hirata:2003ka,Smith:2010gu}. The noise power spectrum of 
the $I$ map, $N_L^{II}$, is given by the shot noise, i.e., the inverse of the total number density per steradian. 
After computing the noise spectra, $N_L^{\grad\grad}$ and $N_L^{II}$, 
the residual B-mode power spectrum \eqref{Eq:delensed-comb} is computed from the above coefficients. 
Note that, if the CMB lensing potential reconstructed from CMB observation is noise dominant 
($\beta_L\to0$), Eq.~\eqref{Eq:delensed-comb} becomes Eq.~\eqref{Eq:resBB:CIB}.

\subsection{Angular power spectrum}

We compute the auto and cross-power spectra of the radio-source number density fluctuations and 
the CMB lensing potential as follows. The angular power spectra between the observables, $X$ and $Y$ 
($X$ and $Y$ are either of $I$ or $\grad$) are described by
\al{
	C_\l^{XY} = \frac{2}{\pi}\Int{}{k}{k^2} P\rom{init}(k) \Delta_\l^X(k)\Delta_\l^Y(k) 
	\,, \label{PS}
}
where $P\rom{init}(k)$ is the scalar power spectrum at an early time with $k$ being the Fourier wave 
number. The functions $\Delta^X_\l(k)$ and $\Delta^Y_\l(k)$ are one of the following (see e.g., 
\cite{Hu:2000ee,LoVerde:2006cj}):
\al{
	\Delta^I_\l(k) &= k^2 \INT{}{z}{}{0}{z_{\rm max}} b(z) \D{N}{z}(z) D(z) j_\l(k\chi(z)) 
	\,, \label{eq:Delta^I} \\
	\Delta^\grad_\l(k) &= \frac{3\Omega\rom{m}H_0^2}{2}
	\notag \\
		&\times \INT{}{\chi}{}{0}{\chi_*} 
			\frac{\chi_*-\chi}{\chi_*\chi}\,\frac{D(z(\chi))}{a(\chi)}\,j_\l(k\chi)
	\,. 
}
The function $j_\l$ is the spherical Bessel function, $a$ is the scale factor, $\chi_*$ is 
the comoving distance to the last scattering surface, $\Omega\rom{m}$ is the fraction of 
the matter energy density, $H\rom{0}$ is the current expansion rate, and $D(z)$ is the growth factor of 
the matter density fluctuations. The quantity $b(z)$ is the halo/galaxy bias, and ${\rm d}N/{\rm d}z$ 
is the source distribution function normalized to unity.

\subsection{CMB experimental specifications}

\begin{table}
\bc
\caption{
CMB experimental specifications used in our analysis, characterized by the following parameters: 
the polarization sensitivity ($\Delta\rom{P}$) in unit of $\mu$K-arcmin, 
beam size ($\theta\rom{FWHM}$) in unit of arcmin, and minimum multipole ($\l\rom{min}$). 
The S3-low and S3-high imply CMB Stage-III (S3) class experiments, 
which will observe before CMB Stage-IV (S4) \cite{Abazajian:2013vfg} will start. 
The S3-wide implies the experiments such as Simons Array and Advanced ACT which will observe nearly 
full sky and be able to provide template of the full-sky lensing-mass map for the \LB
as described in Ref.~\cite{Namikawa:2014yca}. 
}
\label{Table:CMB} \vs{0.5}
\begin{tabular}{lccc} \hline 
 & $\Delta\rom{P}$ & $\theta\rom{FWHM}$ & $\l\rom{min}$ \\ \hline 
Stage-III-low (S3-low)    & 0.5 -- 3.0  & 25.0 & 20  \\ 
Stage-III-high (S3-high)  & 3.0 -- 9.0  & 1.0  & 200 \\ \hline 
\LB                  & 2.0        & 30.0 & 2   \\ 
Stage-III-wide (S3-wide)  & 4.0 -- 12.0 & 4.0  & 200 \\ \hline 
Stage-IV (S4)             & 0.4 -- 1.0  & 3.0  & 2   \\ \hline 
\end{tabular}
\ec
\end{table}

Future ground-based experiments such as an upgrade of BICEP/Keck Array require the delensing analysis 
to suppress the cosmic variance of the lensing B-modes. For efficient delensing, a possible scenario is to 
combine BICEP/Keck with SPT data, because the SPT experiment will observe the same sky with much high 
angular resolution to measure the lensing signals precisely. In future satellite experiments such as 
\LB, to realize efficient delensing, data from high-resolution experiments will be required.
In the case of \LB, the Simons Array \cite{SimonsArray} and Advanced ACT \cite{Calabrese:2014gwa}
are the most likely partners for the delensing analysis. These ground based experiments will observe 
nearly the full sky and will be able to provide a full sky template of the lensing mass map for 
the \LB. Our fiducial analysis assumes that the B-modes observed from these ground based experiments 
are not used for constraining the tensor-to-scalar ratio (though half-wave plates may allow the 
limiting $1/f$ noise to be overcome). Despite this, Ref.~\cite{Namikawa:2014yca}, showed that one can 
measure a full-sky lensing mass map by collecting small patches of the sky to form a nearly full-sky 
patchwork of polarization maps that can be used for the \LB delensing analysis.

For this reason, in our analysis, we assume a delensing analysis for the combination of two CMB experiments, 
where one is a low-resolution high-sensitivity experiment and the other is a high-resolution 
moderate-sensitivity experiment. We vary the sensitivity of the CMB experiments while fixing the angular resolution. 
Their values used in this paper are summarized in Table \ref{Table:CMB}. 
We apply e.g. Eq.~(2.15) of Ref.~\cite{Namikawa:2015tba} to calculate the CMB noise power spectrum.
In addition to these joint delensing analyses, 
we also consider the case with CMB Stage-IV (S4) \cite{Abazajian:2013vfg}.

\subsection{Distribution of SKA mass tracer} \label{subsec:nz}

\begin{figure} 
\bc
\hspace*{-2em}
\includegraphics[width=9cm,clip]{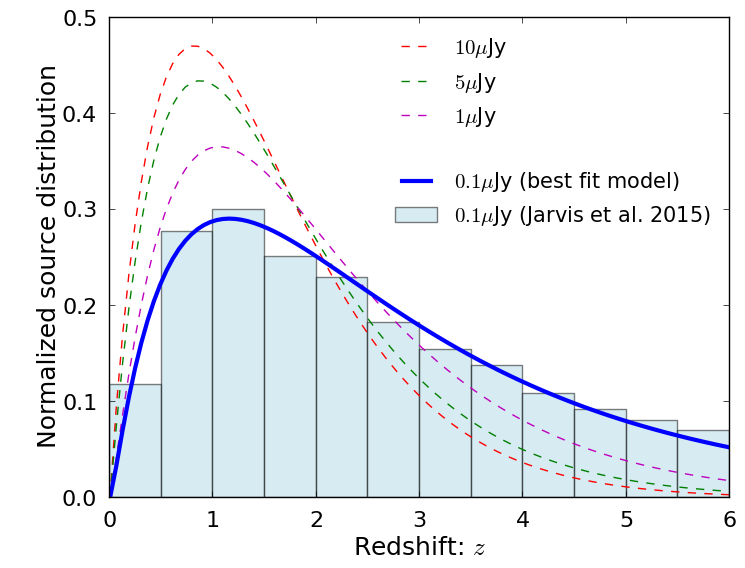}
\caption{
The normalized distribution function of radio sources with the best-fit parameters in four different 
cases of the flux cut ($0.1\mu$Jy, $1\mu$Jy, $5\mu$Jy, and $10\mu$Jy). The histogram of the 
distribution function is taken from Table 1 of Ref.~\cite{Jarvis:2015tqa}.
}
\label{Fig:nz}
\ec
\end{figure}

\begin{table}
\bc
\caption{
The total number of radio sources ($N\rom{tot}$) per deg$^2$ given in Table 1 of 
Ref.~\cite{Jarvis:2015tqa}, the best-fit values of the distribution function parameters 
($p_0$, $p_1$, $p_2$) and the mean redshift ($z\rom{m}$) derived from these best-fit values. The 
functional form of the fitting function is given by Eq.~\eqref{Eq:fz} 
(see Sec.~\ref{subsec:nz} for details)
}
\label{Table:nzfit} \vs{0.5}
\begin{tabular}{l|c|ccc|c} \hline 
Flux cut   & $N\rom{tot}$ & $p_0$ & $p_1$ & $p_2$ & $z\rom{m}$ \\ \hline 
$10\mu$Jy  & $11849$  & 0.92 & 1.04 & 1.11 & 1.53 \\ 
$5\mu$Jy   & $21235$  & 1.01 & 1.14 & 1.02 & 1.66 \\
$1\mu$Jy   & $65128$  & 1.18 & 1.22 & 0.92 & 1.97 \\ 
$0.1\mu$Jy & $183868$ & 1.34 & 1.91 & 0.64 & 2.39 \\ \hline 
\end{tabular}
\ec
\end{table}

We adopt the redshift evolution of the radio sources in Table 1 of Ref.~\cite{Jarvis:2015tqa}, which 
is estimated by the extragalactic simulation of Ref.~\cite{Wilman:2008ew}. We consider the survey with 
four detection thresholds at $1\,{\rm GHz}$ (flux cut): $10\,\mu$Jy, $5\,\mu$Jy, $1\,\mu$Jy and 
$0.1\,\mu$Jy, which are representatives of the surveys conducted with the SKA phase 1 (SKA1) and SKA 
phase 2 (SKA2). The source distribution of the SKA RC survey is given in Table 1 of 
Ref.~\cite{Jarvis:2015tqa}. The total number of the radio sources, $N\rom{tot}$, is $183868$, $65128$, 
$21235$, and $11849$ for $0.1\,\mu$Jy, $1\,\mu$Jy, $5\,\mu$Jy, and $10\,\mu$Jy, respectively. In order 
to have plausible distribution as a function of redshift, we adopt the distribution function with the 
following empirical functional form:
\al{
	\frac{{\rm d}N}{{\rm d}z}(z) \propto z^{p_0} \exp(-p_1z^{p_2}) \,, \label{Eq:fz}
}
where the normalization factor is determined through the condition 
$\INT{}{z}{}{0}{z\rom{max}}{\rm d}N/{\rm d}z=1$. We adopt $z_{\rm max}=6$ as the maximal redshift.
We found that this model provided a good fit to all the relevant redshift distributions with
the four flux cut in terms of the three parameters $(p_0,p_1,p_2)$.
The resultant best-fit values are summarized in Table \ref{Table:nzfit}. 

Fig.~\ref{Fig:nz} shows the normalized distribution function for each flux cut. 
The histogram of the distribution function is taken from Table 1 of Ref.~\cite{Jarvis:2015tqa}, 
while the lines show the above empirical distribution functions with the best-fit values. 
As the flux cut increases, the peak of the distribution shifts to low-redshift. 
To show this clearly, we also show in Table \ref{Table:nzfit} the mean redshift computed
from $z\rom{m}=\INT{}{z}{}{0}{z_{\rm max}} z\,({\rm d}N/{\rm d}z)$.

Since the shape of the distribution function affects the angular power spectra and correlation
coefficient through Eq.~\eqref{eq:Delta^I}, the resultant delensing efficiency
depends on the detection threshold of the survey conducted with the SKA.

\subsection{Bias model of the radio sources}

\begin{table}
\bc
\caption{The best-fit values of the bias parameters.}
\label{Table:bz} \vs{0.5}
\begin{tabular}{l|cccc} \hline 
Flux cut & $b_0$ & $b_1$ & $b_2$ & $b_3$ \\ \hline 
$10\mu$Jy  & -0.0019 & 0.18 & 0.43 & 0.94 \\ 
$5\mu$Jy   & -0.0020 & 0.16 & 0.37 & 0.89 \\
$1\mu$Jy   & -0.0020 & 0.13 & 0.27 & 0.81 \\ 
$0.1\mu$Jy & -0.0019 & 0.11 & 0.20 & 0.76 \\ \hline 
\end{tabular}
\ec
\end{table}

As for the galaxy clustering, namely the biasing, we employ a fit to simulation for the mass function 
${\rm d}n/{\rm d}M$ and the Gaussian linear halo bias factor $b_1$ given in Ref.~\cite{Sheth:1999}.
The weighted averaged bias over the mass range is given by
\al{
	b(z) = \frac{1}{N_{\rm tot}}\INT{}{M}{}{M_{\rm obs}}{\infty}\D{n}{M}(M,z)\,b_1(M,z)  \,,
}
where $N_{\rm tot}=\INT{}{M}{}{M_{\rm obs}}{\infty} {\rm d}n/{\rm d}M$, and $M_{\rm obs}$ is 
the observable mass threshold, which is expected to correspond to the minimum mass of observed radio objects.
We take the value of $M_{\rm obs}$ such that $N_{\rm tot}$ is the total number of radio sources 
for each flux cut given in Table 1 of Ref.~\cite{Wilman:2008ew}.

Since the bias factor has uncertainties, several nuisance parameters would be included in cosmological 
analysis. We parametrize the redshift evolution of the bias with the third order polynomial function: 
$b(z)=b_3+b_2z+b_1z^2+b_0z^3$. We checked that the bias model we used so far is well fitted by this 
third order polynomial, and we use the best-fit values as the fiducial values of $b_i$. The best-fit 
values of the bias parameters are summarized in Table \ref{Table:bz}. Hereafter we will consider not 
only $(p_0,p_1,p_2)$ but also $(b_0,b_1,b_2,b_3)$ as the free parameters to quantify the impact of 
uncertainties of the bias model and source distribution on the parameter estimation. 
Note that the non-linear bias also becomes important especially for small scales and low-redshift, 
though the delensing efficiency is sensitive to the multipoles at $L<500$ and mass tracers at high 
redshifts. A thorough investigation of the impact of the non-linear bias requires realistic simulations 
for the SKA observation, and this effect will be addressed in our future work. 

\section{Results} \label{results}

In this section, we present the results of the expected delensing performance by adding 
data from the SKA RC survey. We then show the impact of the possible uncertainties 
on the sensitivity to the tensor-to-scalar ratio. 
We also discuss comparisons between our results and other methods. 

\subsection{Correlation between CMB lensing and galaxy} 

\begin{figure} 
\bc
\hspace*{-1em}
\includegraphics[width=8.8cm,clip]{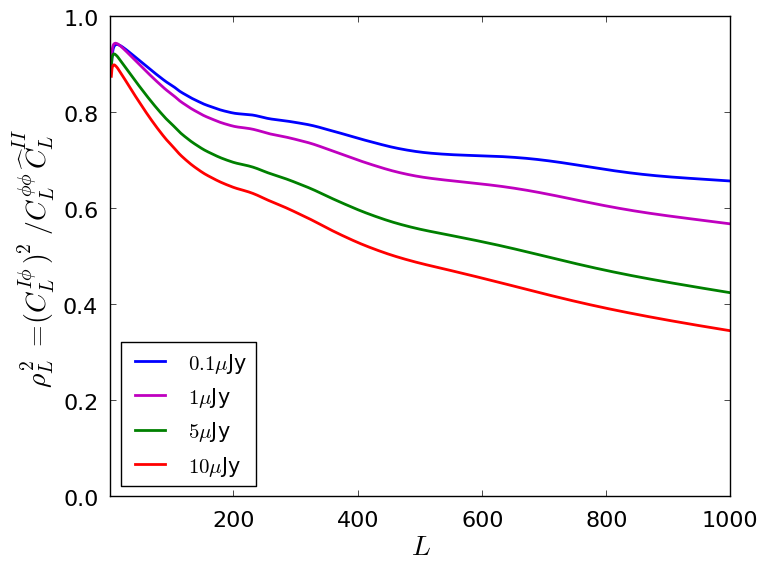}
\caption{
Expected correlation coefficient for four different cases of the flux cut 
($0.1\mu$Jy, $1\mu$Jy, $5\mu$Jy and $10\mu$Jy).
}
\label{Fig:corr}
\ec
\end{figure}

\begin{figure*} 
\bc
\hspace*{-1em}
\includegraphics[width=8.8cm,clip]{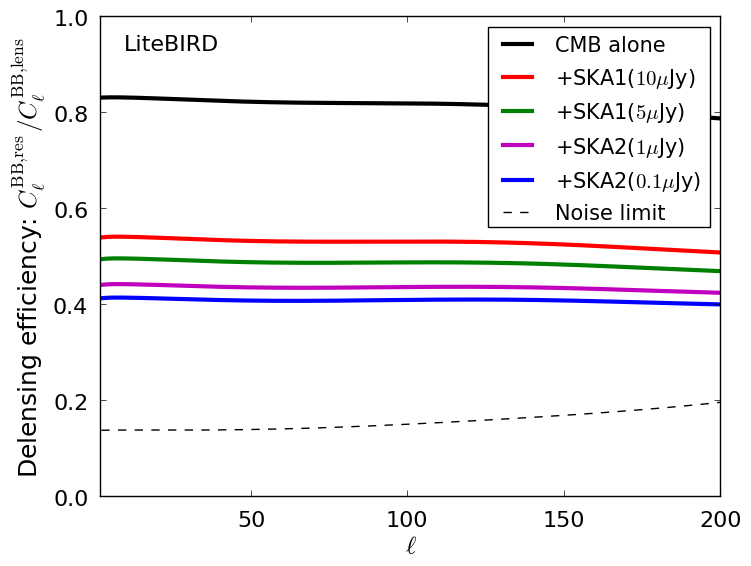}
\includegraphics[width=8.8cm,clip]{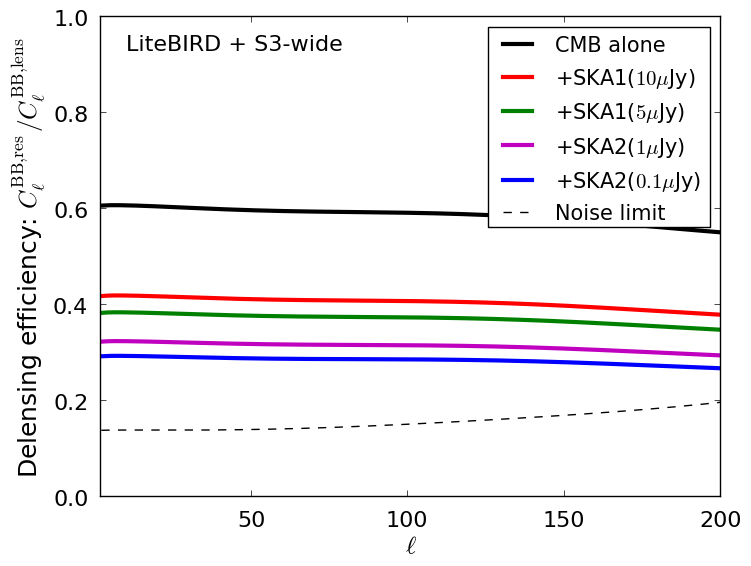}
\caption{
The delensing efficiency for the joint analysis of \LB and SKA (Left) and of \LB, S3-wide and 
SKA (Right). Here, the polarization sensitivity of S3-wide is $6\mu$K-arcmin. The lines show the case 
with the CMB data alone, CMB+SKA1 ($10\mu$Jy), CMB+SAK1 ($5\mu$Jy), CMB+SKA2 ($1\mu$Jy) and 
CMB+SKA2 ($0.1\mu$Jy), respectively. The thin dashed line shows the limit due to the presence of 
the CMB instrumental noise in B-modes. 
}
\label{Fig:delensing}
\ec
\end{figure*}

Fig.~\ref{Fig:corr} shows the correlation coefficient defined in Eq.~\eqref{Eq:corr}. The noise power 
spectrum is given by the shot noise derived from the source number density per steradian. As the flux 
cut increases, the minimum mass of the observed radio sources increases, implying that the values of 
the bias parameters become large. On the other hand, the mean redshift of the source distributions 
decreases as the flux cut increases. The shot noise term in $\hC^{II}_L$ also increases as the flux 
cut increases because the total number of radio sources decreases. From these effects, the correlation 
coefficient becomes small for larger flux cut

Fig.~\ref{Fig:delensing} shows the delensing efficiency, i.e., the ratio of the angular power spectrum of 
the residual B-modes to that of the lensing B-modes; $C_\l^{\rm BB,res}/C_\l^{\rm BB,lens}$.
We show the cases for the joint analysis between (1) \LB for the CMB observations
and the SKA as an external data of the mass tracer and (2) \LB, S3-wide and SKA observations.
During the \LB observation, the SKA2 will start their observation, and 
therefore we consider the joint delensing analysis with not only the SKA1 but also the SKA2. 
Note that the delensing efficiency is computed from the correlation coefficients in Eq.~\eqref{Eq:corr}.
With the CMB data alone, the power spectrum amplitudes of the residual lensing B-modes become 
$\sim 60$\% of original lensing B-mode power-spectrum amplitudes. 
Combining the SKA2 data, the delensing analysis removes $\sim70$\% of the lensing B-modes 
in the measured B-mode power spectrum.

Note that, in the above analysis, we fix the specific model for the mass function. 
We check whether the use of different mass functions changes the delensing efficiency, 
finding that the changes are negligible with the Press-Shechter mass function 
\cite{PS:1974} or MICE \cite{Crocce:2010}. 

\subsection{Improvement to constraints on tensor-to-scalar ratio} 

\begin{figure} 
\bc
\hspace*{-2em}
\includegraphics[width=8.8cm,clip]{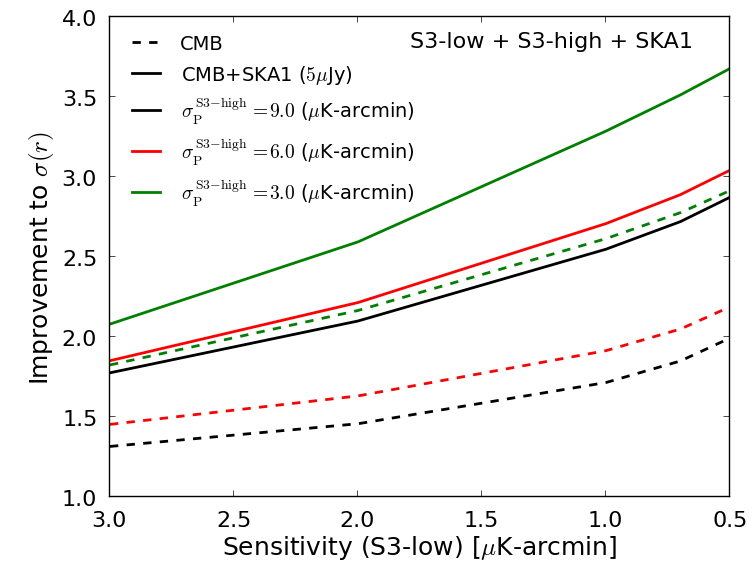}
\caption{
The expected improvement to the constraints on the tensor-to-scalar ratio, $\alpha$, using 
CMB observations alone or combining the SKA1. We combine two different CMB experiments 
where one has high-sensitivity but low angular resolution (S3-low)
and the other has low-sensitivity and high-angular resolution (S3-high).
}
\label{Fig:BKSPT}
\ec
\end{figure}

\begin{figure} 
\bc
\hspace*{-2em}
\includegraphics[width=8.8cm,clip]{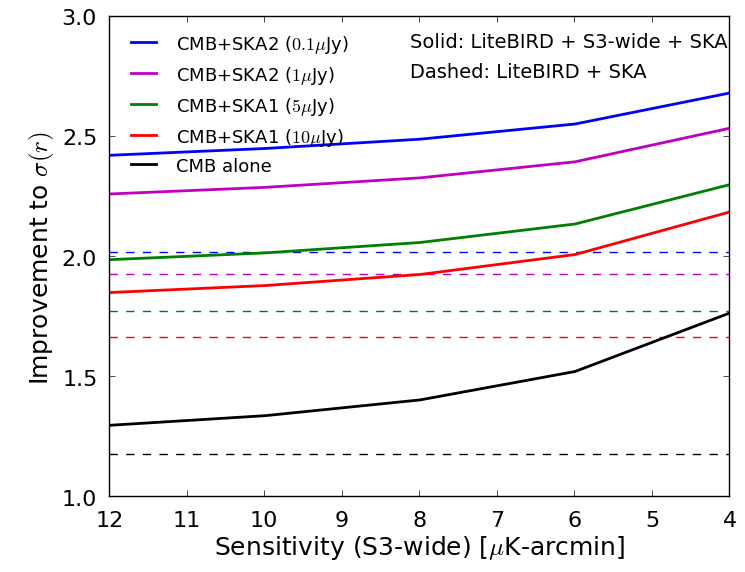}
\caption{
The expected improvement to the constraints on the tensor-to-scalar ratio 
for the joint analysis between the \LB, S3-wide and SKA (thick solid lines). 
We also show the cases with the \LB and SKA (thin dashed lines). 
}
\label{Fig:LBSA}
\ec
\end{figure}

\begin{figure} 
\bc
\hspace*{-2em}
\includegraphics[width=8.8cm,clip]{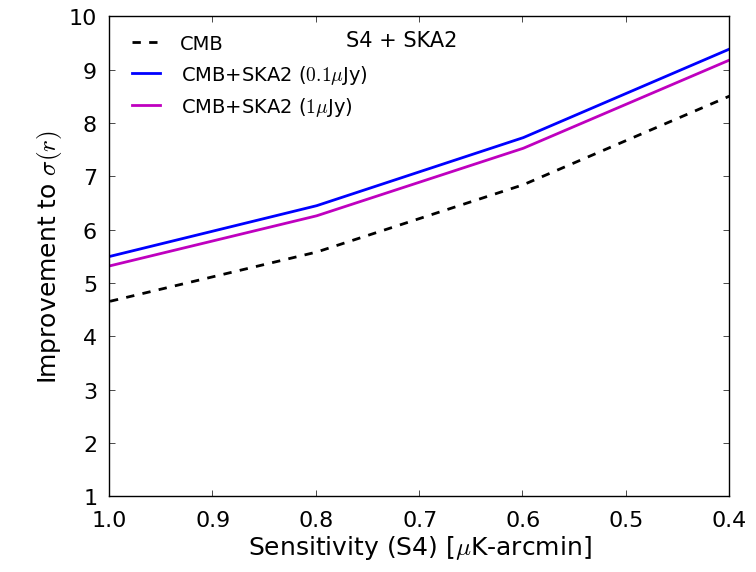}
\caption{
Same as Fig.~\ref{Fig:LBSA} but for the joint analysis between the S4 and SKA2.
}
\label{Fig:S4}
\ec
\end{figure}

We will now discuss the improvement to the constraint on the tensor-to-scalar ratio, $\sigma(r)$. 
Since the lensing and residual B-modes are flat at $\l<100$, the inprovement to the constraints on 
the tensor-to-scalar ratio (ratio of the constraints with the delensing to that without the delensing) 
is approximately given by (see e.g. \cite{Smith:2010gu,Sherwin:2015baa}): 
\al{
	\alpha \equiv \AVE{\frac{C^{\rm BB,lens}_{\l}+N^{\rm BB}_{\l}}{C^{\rm BB,res}_{\l}+N^{\rm BB}_{\l}}}
	\,.
}
Here $\ave{\cdots}$ is the averaged value between $\l=\l\rom{min}$ and $100$, and $N^{\rm BB}_\l$ is 
the B-mode noise power spectrum. The cosmic variance of the primordial B-modes is ignored in the above 
equation (i.e., the fiducial value of the tensor-to-scalar ratio is $r=0$). 

During the SKA1 survey, the near future CMB experiments such as significant upgrades of BICEP/Keck Array 
and SPT plan to observe the B-modes on large angular scales with a high polarization sensitivity. 
These experiments will realize an efficient delensing analysis. Even in this case, the delensing with 
the SKA1 data is expected to help the removal of the lensing B-modes. As discussed in previous section, 
we characterize these experiments as the S3-low and S3-high described in Table \ref{Table:CMB}.

In Fig.~\ref{Fig:BKSPT} we show the expected improvement to the constraints on the tensor-to-scalar 
ratio, $\alpha$, for the joint analysis of the S3-low, S3-high and SKA1. Combined with CMB and SKA1 
data, delensing will improve the constraint on the tensor-to-scalar ratio by more than a factor of 
$\sim 2$ depending on the polarization sensitivities of the two CMB experiments. Compared to the CMB 
delensing alone, the delensing analysis with the SKA1 data will further improve the sensitivity to the 
tensor-to-scalar ratio by a factor $1.2$--$1.6$. For the high-sensitivity cases, the delensing with 
the CMB data alone will significantly suppress the lensing B-modes, but the delensing improvement from 
the SKA1 data would be still non-negligible. 

In Fig.~\ref{Fig:LBSA}, we show $\alpha$ for the joint analysis of \LB, S3-wide, and SKA. In this case, 
delensing with the SKA data will play an important role in probing the primary B-modes. Combining CMB 
and SKA2 data, the improvement becomes a factor of $2.3$--$2.7$, while using the CMB data alone 
improves the constraints by $1.3$--$1.7$. In our fiducial analysis, the minimum multipole of the E- 
and B-modes from the S3-wide experiment is $\l\rom{min}=200$, but we checked that the inclusion of the 
lower multipoles improve within $\leq 5$\% at $\Delta\rom{P}=6\mu$K-arcmin. This is because the B-mode 
noise is mostly determined by the \LB polarization noise, and also because the lensing reconstruction 
noise is not sensitive to the inclusion of the large-scale E- and B-modes. Measurements of the 
large-scale polarization from the S3-wide are, therefore, not very important for the \LB delensing. 
Note that the lensing B-modes are removed significantly even using the SKA1 data. Note also that the 
reason why the SKA helps the delensing is explained as follows: As discussed in 
Ref.~\cite{Smith:2010gu}, the contributions to the lensing B-modes at large scales ($\l<200$) come 
from the CMB lensing potential at $30\alt\l\alt1000$. The lensing signals reconstructed from the \LB 
and S3-wide are, however, noisy at smaller scales. Although the SKA (and also Planck CIB) mass fields 
are not perfectly correlated with the CMB lensing potential at all scales, they are correlated with 
the CMB lensing potential even at smaller scales (see Fig.~\ref{Fig:corr}). Therefore, delensing 
combined with the SKA, especially for lower flux cuts, further improves the efficiency. 

Fig.~\ref{Fig:S4} shows the case with the S4 experiment combined with the SKA2. If the S4 experiment 
is realized, the noise level in the CMB polarization map will be less than $1\mu$-arcmin with high 
resolution ($\leq 3$ arcmin), and approximately $80$\%--$90$\% of the lensing B-modes will be removed 
using the CMB data alone. As discussed in previous work, \cite{Simard:2014aqa,Sherwin:2015baa}, the 
delensing analysis with the S4 data alone is more efficient than that with the CIB data. If we 
additionally use the SKA data, however, the constraints on $r$ would be further improved by 
a non-negligible amount compared to case of internal CMB delensing alone (with $1\mu$K-arcmin).

\subsection{Uncertainties of the bias model and source distribution} 

\begin{table}
\bc
\caption{
The fractional degradation to the $\sigma(r)$ due to the bias model and source distribution 
uncertainties, i.e., the constraints on $r$ by additionally marginalizing the bias model and source 
distribution parameters divided by that without the marginalization of these parameters. The 
polarization sensitivity of the S3-low, S3-high and S3-wide are $1\mu$K-arcmin, $6\mu$-arcmin and 
$6\mu$K-arcmin, respectively. The flux cuts of the SKA1 and SKA2 RC surveys are $5\mu$Jy and 
$0.1\mu$Jy, respectively.
}
\label{Table:fisher} \vs{0.5}
\begin{tabular}{l|cc} \hline 
 & $B^{\rm res}$ & $B^{\rm res}+\grad+I$ \\ 
\hline 
S3-low + S3-high + SKA1   & 10.4 & $3.82\times 10^{-3}$ \\ 
\LB + S3-wide + SKA2 & 3.85 & $3.63\times 10^{-3}$ \\
\hline
\end{tabular}
\ec
\end{table}

In a realistic analysis, the bias model should be determined with other information such as 
$C_\l^{\grad I}$ and $C_\l^{II}$. The source distribution will be also determined by other SKA survey 
plans and follow-up observations in other wavelength, but the bias parameters and redshift distribution 
are not completely determined by these follow-up observations. In order to severely determine the 
parameters ($b_i$ and $p_i$), in actual analysis, we marginalize these parameters simultaneously in 
addition to $r$. We discuss here how the additional marginalization of the bias model and source 
distribution parameters degrade the constraints on $r$. 

To take into account the uncertainties from the bias model and source distribution function, 
we perform the Fisher matrix analysis to obtain the expected constraints on the tensor-to-scalar ratio 
by marginalizing the tensor-to-scalar ratio, $r$, the bias parameters, $b_i$ ($i=0,1,2,3$), 
and the parameters of the source distribution, $p_i$ ($i=0,1,2$). We define the Fisher matrix as
\al{
	F_{ij} &\equiv \sum_{\l=\l\rom{min}}^{100} \frac{2\l+1}{2[C_\l^{\rm BB,res}+N_\l^{\rm BB}]^2}
		\PD{C^{\rm BB,res}_\l}{\theta_i}\PD{C^{\rm BB,res}_\l}{\theta_j}
	\notag \\
	&+ \sum_{L=2}^{1000} \frac{2L+1}{2} 
		\Tr \left[\bR{C}^{-1}_L\PD{\bR{C}_L}{\theta_i}\bR{C}^{-1}_L\PD{\bR{C}_L}{\theta_j}\right]
	\,, 
}
where $\theta_i=r$, $b_i$ or $p_i$ \footnote{In our analysis, since the fiducial value of $r$ is zero, 
the derivative of $C_\l^{\rm BB,res}$ is non-zero if $\theta_i=r$.}. 
The covariance matrix of the mass tracers $\bR{C}_L$ is given in Eq.~\eqref{Eq:cov}. 
We compute the derivatives of the power spectra as described in appendix \ref{sec:deriv}. 
Note that we omit the correction of the partial sky coverage since it does not affect our results, i.e., 
we only focus on the ratio of the constraints, and assume that CMB and SKA data sets are taken only 
from their overlapped region. We do not include the non-Gaussian covariance of the residual B-mode 
power spectrum since it is negligible \cite{Namikawa:2015tba}. 

In our calculation, we assume that the same $\bm{a}_L$ is used for computing theoretical predictions 
and delensed B-modes, and $\bm{a}_L$ is fixed in the parameter estimation. For this reason, we do 
not consider the dependence of $\bm{a}_L$ on $b_i$ and $p_i$ in $\pd C^{\rm BB,res}/\pd \theta_i$. 
Then, the uncertaintes in $\bm{a}_L$ could make the delensed B-modes suboptimal. We however ignore 
the uncertainties in $\bm{a}_L$ given in Eq.~\eqref{Eq:filter-comb} because the variance of the 
residual B-modes ($C_\l^{\rm BB,res}$) is insensitive to the uncertainties in $\bm{a}_L$. To see 
this, let us consider the case where $\bm{a}_L$ is given by the true $\bm{a}_L$ 
($\bm{a}_L^{\rm true}$) with a small correction ($\delta \bm{a}_L$) i.e. 
$\bm{a}_L=\bm{a}_L^{\rm true}+\delta\bm{a}_L$. Substituting this into Eq.~\eqref{Eq:delensed-ClBB}, 
however, $C_\l^{\rm BB,res}$ contains only higher-order terms of $\delta \bm{a}_L$. This is because 
the coefficients are derived so that $\bm{a}_L$ satisfies ${\rm d}C_\l^{\rm BB,res}/{\rm d}\bm{a}_L=0$ 
around $\bm{a}_L=\bm{a}_L^{\rm true}$, i.e., the first order Taylor expansion in terms of $\bm{a}_L$ 
around $\bm{a}_L^{\rm true}$ vanishes. 

In Table \ref{Table:fisher}, we summarize the degradation factor. Using the residual B-modes alone 
significantly degrades the sensitivity to the bias model and source distribution uncertainties. 
However, inclusion of the mass tracers will strongly constrain these parameters, and the degradation 
due to these uncertainties becomes negligible. We also checked the cases with other polarization 
sensitivities and flux cut given in Table \ref{Table:CMB}, but the values are almost unchanged. 
Note that $b_i$ and $p_i$ are severely constrained by the mass-tracer power spectrum obtained from 
the SKA survey. The $1\sigma$ uncertainties of $b_i$ are within percent level, while those of $p_i$ 
are approximately between few to ten percents.

\subsection{Comparison with alternative delensing methods} 

We will now discuss the delensing efficiency presented in this paper in comparison with that of other 
delensing methods. 

In the SKA survey, it is also possible to use the intensity mapping of the 21cm fluctuations 
generated by high-z sources. The lensing mass fields of the 21cm fluctuations in the intensity map 
can be reconstructed using the same methodology of the CMB lensing reconstruction \cite{Zahn:2006}, 
or by measuring the shape of each galaxy \cite{Marian:2007sr}. Ref.~\cite{Sigurdson:2005cp} showed 
that reconstructed lensing mass fields from futuristic 21cm surveys is useful for the delensing 
analysis. However, this method requires high-redshift sources ($z\rom{s}=\mC{O}(10)$), and suffers 
from contamination by foreground emission. 

The CIB can also be used as a lensing mass tracer, and is known to be a possible candidate for 
the delensing analysis in near future CMB experiments \cite{Simard:2014aqa,Sherwin:2015baa}. 
According to Fig.~1 of Ref.~\cite{Sherwin:2015baa}, the correlation coefficient, $\rho_L^2$, of 
the SKA2 (SKA1) is comparable to (smaller than) that of the Planck CIB observation.
While SKA1 data is hence less efficient at delensing, its addition may be useful for further increasing 
delensing performance. Furthermore, dust foregrounds in the CIB measurements may significantly reduce 
the correlation coefficient on large scales. Although the impact of foreground residuals can be 
suppressed by filtering out low multipoles of the CIB fluctuations, this can cause some loss of 
delensing performance, especially if high-dust regions are included in the analysis. Finally, 
uncertainties in the level of foreground residuals may be problematic. Therefore, the use of the SKA1 
data could be of great assistance to CIB delensing, complementing the CIB data on scales where dust 
foregrounds may be large and allowing for cross-checks.

\section{Summary} \label{summary}

We have discussed the potential use of the SKA RC survey for delensing future CMB experiments such as 
a significant upgrade of BICEP/Keck Array and \LB. We found that joint delensing using near future CMB 
experiments and the SKA1 survey will improve the constraints on the tensor-to-scalar ratio 
significantly (by more than a factor of $\sim 2$) compared to those without the delensing analysis. 
Compared to the use of CMB data alone, the inclusion of the SKA1 data will increase the significance of 
the constraints on the tensor-to-scalar ratio by a factor $1.2$--$1.6$. 
We also explored the case of a joint analysis of \LB, a wide-field 
CMB experiment, and the SKA2, finding that the lensing B-modes will be significantly reduced by delensing.
The constraints will be improved by a factor of $2.3$--$2.7$ compared to that without delensing.
In particular, compared to the delensing with the CMB data alone, the inclusion of the SKA2 data 
further improves the constraints by approximately a factor of $2$. We then discussed the impact of 
the uncertainties in the galaxy bias and source distribution on $\sigma(r)$ based on a Fisher matrix 
analysis, showing that the impact of these uncertainties is negligible because the parameters 
associated with the bias model and source distribution are be strongly constrained by information on 
the mass tracers. The map from the SKA RC survey will, therefore, be quite useful for future CMB 
delensing analyses, especially for \LB observations.

In this paper, we have made several assumptions. For example, we have assumed that the extragalactic 
sky simulation of Ref.~\cite{Wilman:2008ew} provides plausible estimates of the redshift distributions 
of source populations, and that any residual foregrounds in the radio source maps are negligible.
Although the simulation is so far in good agreement with latest radio observations~\cite{Condon:2012ug},
the above assumptions may not be valid for future radio surveys on large scales. We also have assumed 
that the radio source maps obey Gaussian statistics. Intrinsic non-Gaussianity in the radio source 
distributions would, however, bias the amplitude of the residual B-mode power spectrum and its error.
These issues should be addressed with an appropriate extragalactic sky simulations, and are left for 
future work.

\begin{acknowledgments}
T.N. is supported by Japan Society for the Promotion of Science (JSPS) fellowship for abroad (No.~26-142). 
D.Y. is supported by Grant-in-Aid for JSPS Fellows (No.~259800). 
We would like to thank Masamune Oguri for useful comments. 
\end{acknowledgments}

\appendix

\section{Derivative} \label{sec:deriv}

In this appendix, we show the derivative of the residual B-mode power spectrum with respect to 
the bias and distribution function parameters, $b_i$ and $p_i$. 

The filter function is fixed in a realistic analysis. If the observed quantities, $C_L^{I\grad}$ and 
$C_L^{II}$, are changed as 
\al{
	C_L^{I\grad} &\to C_L^{I\grad}(1+\Delta^{I\grad}_L) \,, \notag \\ 
	C_L^{II} &\to C_L^{II}(1+\Delta^{II}_L) \,, 
}
the residual B-mode becomes \cite{Sherwin:2015baa}
\al{
	\Delta C^{\rm BB,res}_\l &= \Xi^{\l'L}_\l C^{\rm EE}_{\l'}\Cgg_LW_{\l'}\rho^2_L
	\notag \\
	&\qquad \times \bigg[-2\Delta^{I\grad}_L+\frac{C_L^{II}}{C_L^{II}+N_L^{II}}\Delta^{II}_L \bigg]
	\,,
}
where $\rho^2_L$ is computed with the fiducial power spectrum. The derivative with respect to a 
parameter $\theta_i$ is then given by
\al{
	&\PD{C^{\rm BB,res}_\l}{\theta_i} = \Xi^{\l'L}_\l C^{\rm EE}_{\l'}\Cgg_LW_{\l'}\rho^2_L
	\notag \\
	&\qquad \times \bigg[-2\PD{\ln C^{I\grad}_L}{\theta_i}+\frac{1}{C_L^{II}+N_L^{II}}\PD{C^{II}_L}{\theta_i} \bigg]
	\,, \label{Eq:res-RC}
}

For the delensing combined with the CMB and SKA observations, we obtain
\al{
	\Delta C^{\rm BB,res}_\l 
		&= \Xi^{\l'L}_\l W_{\l'}C^{\rm EE}_{\l'} \bigg[2(-1+a^{\grad}_L)a^I_L \Delta C^{I\grad}_L
	\notag \\
		&\qquad + (a^I_L)^2 \Delta C^{II}_L \bigg]
	\,,
}
where the filter functions are given by (see Eq.~\eqref{Eq:filter-comb})
\al{
	\binom{a_L^{\grad}}{a^I_L} 
		= \frac{1}{1-\beta_L\rho^2_L}\binom{\beta_L(1-\rho^2_L)}{\gamma_L(1-\beta_L)}
	\,,
}
and the quantitites, $\beta_L$ and $\gamma_L$, are defined in Eqs.~\eqref{Eq:beta} and 
\eqref{Eq:gamma}, respectively. This leads to
\al{
	&\PD{C^{\rm BB,res}_\l}{\theta_i} 
		= \Xi^{\l'L}_\l C^{\rm EE}_{\l'} \Cgg_LW_{\l'}\rho^2_L\frac{(1-\beta_L)^2}{(1-\beta_L\rho^2_L)^2}
	\notag \\
	&\qquad \times \bigg[-2\PD{\ln C^{I\grad}_L}{\theta_i}
		+\frac{1}{C_L^{II}+N_L^{II}}\PD{C^{II}_L}{\theta_i} \bigg]
	\,.
}
If the CMB lensing potential reconstructed from CMB observation is noise dominant ($\beta_L\to0$), 
the above equation becomes Eq.~\eqref{Eq:res-RC}. The above equation is used to evaluate the 
derivatives of the residual B-mode power spectrum in our Fisher matrix analysis. 

\bibliographystyle{apsrev}
\bibliography{cite}

\end{document}